# Research on Core Loss of Direct-drive 75kW Tidal Current Generator Using Machine Learning and Multi-objective Optimization Algorithms

Shuai Zu[1,2], Wanqiang Zhu[1,2*], Fuli Zhang[1,2], Chi Xiao[1,2], Xiao Zhang[3], Yixiao Li[1,2], Xinze Wen[1,2], Yingying Qiao[1,2], Junyi Xu[1,2]

1. *School of Physics, Northeast Normal University, 5268 Renmin Street, Changchun, Jilin Province, People's Republic of China*
2. *JiLin Provincial Key Laboratory of Advanced Energy Development and Application Innovation, 5268 Renmin Street, Changchun, Jilin Province, People's Republic of China*
3. *Changchun institute of technology*
4. *\* Correspondence: Zhuwq773@nenu.edu.cn*

**Abstract:** This paper presents a classification of generator excitation waveforms using principal component analysis (PCA) and machine learning models, including logistic regression, random forest, and gradient boosting decision trees (GBDT). Building upon the traditional Steinmetz equation, a temperature correction term is introduced. Through nonlinear regression and least squares parameter fitting, a novel temperature correction equation is proposed, which significantly reduces the prediction error for core losses under high-temperature conditions. The average relative error is decreased to 16.03%, thereby markedly enhancing the accuracy. Using GBDT and random forest regression models, the independent and combined effects of temperature, excitation waveforms, and magnetic materials on core loss are analyzed. The results indicate that the excitation waveform has the most significant impact, followed by temperature, while the magnetic core material exhibits the least influence. The optimal combination for minimizing core loss is identified, achieving a core loss value of 35,310.9988 under the specified conditions. A data-driven predictive model for core loss is developed, demonstrating excellent performance with an R² value of 0.9703 through machine learning regression analysis, indicating high prediction accuracy and broad applicability. Furthermore, a multi-objective optimization model considering both core loss and transmission magnetic energy is proposed. Genetic algorithms are employed for optimization, resulting in an optimal design scheme that minimizes core loss and maximizes transmission magnetic energy. Based on this model, the magnetic core compensation structure of a direct-drive 75kW tidal energy generator is optimized in practical applications, yielding satisfactory results.

**Keywords:** Ocean Energy, magnetic core loss, Steinmetz equation, machine learning, genetic algorithm, optimization model

# 1. Introduction

In the contemporary era of rapid scientific and technological advancement, marine energy, as a clean and sustainable resource, has garnered widespread attention for its development and utilization[1]. As a critical component in converting oceanic energy into electrical power, the optimization of the magnetic core plays a pivotal role in enhancing the efficiency, stability, and reliability of the entire power generation system[2]. Magnetic core losses in ocean energy generators not only lead to energy wastage but also compromise generator performance, shorten operational lifespan, and potentially jeopardize the stability of the entire marine energy power plant[3]. Consequently, an in-depth investigation into magnetic core losses and the pursuit of effective analysis and improvement methods have become focal points within both academic and industrial circles in the field of marine energy[4].

The study of magnetic core losses in ocean energy generators holds significant importance[5]. From a strategic perspective on energy utilization, the efficient harnessing of ocean energy becomes increasingly crucial as global demand for clean energy continues to rise[6]. Reducing magnetic core losses can significantly enhance the conversion efficiency of ocean energy, minimize energy waste, and contribute to achieving sustainable energy development goals[7]. For ocean energy generation equipment, minimizing magnetic core losses ensures more stable generator operation, reduces failures due to overheating, extends equipment lifespan, and lowers offshore maintenance costs—critical factors for devices operating in complex marine environments. Optimizing magnetic core performance can improve power transmission and conversion efficiency, reduce energy consumption, and enhance system reliability and stability under harsh ocean conditions[8].

Substantial progress has been made in the research on magnetic core losses in marine energy generators internationally. By optimizing material composition and microstructure, U.S. research teams have successfully reduced magnetic core losses, particularly in low-frequency applications in deep-sea environments, through the adoption of advanced nanocrystalline material synthesis technologies[9]. German researchers have focused on heat dissipation design, addressing the high humidity and corrosive nature of marine environments by improving packaging structures and cooling methods, thereby effectively mitigating temperature-induced losses[9]. Japanese researchers have achieved important breakthroughs in modeling and predicting magnetic core losses using sophisticated physical models and numerical simulations combined with marine environmental parameters to accurately forecast losses under various sea states and operating conditions[11].

In China, research on magnetic core losses in ocean energy generators is also advancing. Several universities and research institutions have made strides in developing magnetic core materials, enhancing their magnetic properties and loss resistance in marine environments through modifications and optimizations of traditional materials[12]. Domestic enterprises are actively exploring ways to reduce magnetic core losses by refining production processes and optimizing product designs, thereby improving the quality and performance of marine energy generator cores for complex marine conditions[13]. However, despite notable achievements both domestically and internationally, challenges remain. The current magnetic core loss models for ocean energy generators are insufficiently robust to accurately describe loss behavior under complex marine environments and variable operating conditions. Additionally, the application of multi-objective optimization in analyzing magnetic core losses requires further exploration and innovation to meet the demands of efficient and stable operation of ocean energy power generation systems.

With the rapid development and broad application of machine learning technology, significant advantages have emerged in processing complex data and building predictive models[14]. By analyzing extensive datasets from ocean energy generator operations, machine learning models can uncover hidden patterns and relationships, providing new perspectives and methodologies for analyzing magnetic core losses. Moreover, advancements in multi-objective optimization algorithms enable finding optimal solutions among conflicting objectives[15], facilitating more efficient and reliable operation of ocean energy power generation systems.

Table 1 Symbol description1

| Symbols | Meaning | Units |
| --- | --- | --- |
| D | Training set | \ |
| d | Euclidean distance | \ |
| w | Weights | \ |
| F1 | F1 scores | \ |
| N | Type of core material | \ |
| B | Peak magnetic flux density | |
| F | Excitation waveform | \ |
| f | Frequency | Hz |
| T | Temperature | Degrees Celsius |
| X | Eigenvalues | |
| P | Core loss | \ |
| Alpha. | Temperature correction factor | $W/m^3$ |
| L | Loss function | \ |
| ME | Transmission magnetic energy | \ |
| | | J |

## 2. Research and analysis of influencing factors of magnetic core loss

In the field of magnetic core loss analysis and model improvement of ocean energy generators, in order to deeply analyze the independent influence of temperature, excitation waveform and magnetic core material on magnetic core loss and the synergistic mechanism, this study introduced GBDT (gradient lifting decision Tree)[16] and random forest algorithm[17] in machine learning to carry out regression analysis of relevant data. Through the powerful feature learning and data processing capabilities of these two algorithms, the complex nonlinear relationship between various factors and magnetic core loss can be accurately analyzed, so as to achieve effective prediction and model optimization of magnetic core loss, and provide solid data support and theoretical basis for efficient and stable operation of ocean energy generators[18].

## 2.1 Determination of targets and features

The correlation regression model is used to find the mapping relationship f(X) between input feature X and target value y, and to evaluate the independent and synergistic effects of each feature on core loss y. For the features, we input the data set:

$$D = \{(X_i, y_i)\}_{i=1}^{n} \qquad (1)$$

Where: represents the three characteristics of each sample i: $X_i = (X_{i1}, X_{i2}, X_{i3})$

$$\begin{cases} X_{i1} : \text{core material} \\ X_{i2} : \text{temperature} \\ X_{i3} : \text{Excitation waveform} \end{cases} \qquad (2)$$

## 2.2 Selection and establishment of regression model

In order to enhance the stability of the model, the number of control decision trees is 100; In order to control the complexity of the tree and prevent overfitting, we set the maximum depth of each tree to 10; The minimum number of samples for leaf nodes is 1, and the minimum number of samples for internal node splits is 2; And the maximum number of leaves is 100; For the decision tree, the number of index features randomly extracted each time is set to 4. Basic parameter Settings are shown in the following table:

Table 2 Setting of basic parameters of random forest

| Basic parameters | Numerical values |
| --- | --- |
| Number of decision trees | 100 |
| Maximum depth of decision tree | 10 |
| Maximum number of leaf nodes | 50 |
| Minimum number of leaf nodes | 1 |
| Minimum number of samples for internal node classification | 2 |
| Index characteristics of node splitting extraction | 4 |

**(1) Construction of decision tree**

This model uses Bootstrap sampling (random sampling with put back) to generate different training data sets, thereby increasing the diversity of the model. For each tree, after given training set *D*, n samples are randomly selected to form a new training set $D_b$:

$$\begin{cases} D = \{(x_i, y_i)\}_{i=1}^{n} \\ D_b = \{(x_i', y_i')\}_{i=1}^{n} \end{cases} \qquad (3)$$

Where: It is obtained from the original training set by placing back the samples. $x_i'$

At each node splitting, 4 nodes are randomly selected from the enterprise's reputation rating index for splitting, and the optimal splitting point among these features is found out.

Since it is necessary to classify each data, the optimal split point is selected according to the recursive method of information gain, so as to construct the decision tree:

$$\text{Gain}(D,k) = H(D) - \sum_{v \in \text{Values}(k)} \frac{|D_v|}{|D|} H(D_v) \tag{4}$$

Where: is the entropy of data set D; $H(D)$ $D_v$ is the subset of feature $k$ whose value is v.

When the depth of splitting to the tree reaches 10 or the number of node samples after splitting is less than 2, the node splitting is stopped. At the same time, when training the suitability of each decision tree, the gain values in all trees are added up to get the importance of this feature[19]. The calculation formula is as follows:

$$\text{FeatureImportance}(X_k) = \frac{1}{n_{\text{trees}}} \sum_{i=1}^{n_{\text{trees}}} \sum_{k \in \text{nodes}} \Delta \text{Impurity}(k) \cdot \mathbb{I}(X_j \text{ is used in } k) \tag{0.0.1}$$

At the same time, the important value of each index feature is divided by the importance of all features, so as to achieve the standardization of the importance of each feature indicator, so that the sum of the importance of all feature values is 1:

$$\sum_{k=1}^{8} \frac{\text{FeatureImportance}(X_k)}{\sum_{k=1}^{8} \text{Feature Importance}(X_k)} = 1 \tag{5}$$

According to the above formula, the quality pairs of feature importance obtained are sorted from high to low, so as to determine the index that has the most important influence on the model prediction.

(2) **The integration of multiple decision trees**

In order to improve the accuracy and stability of the model, and make the evaluation results of each enterprise's reputation grade more reliable, the prediction results of multiple decision trees are adopted, and the MajorityVoting method is adopted to determine the final prediction results:

$$\hat{y} = \text{mode}(\{T_i(x)\}_{i=1}^{n_{\text{trees}}}) \tag{6}$$

$T_i(x)$ is *the* prediction result of the i th tree for sample *x*.

In order to improve the decision tree, we introduce a new decision tree -- the core idea of *GBDT* is to treat the construction of the model as an iterative optimization process and improve the performance of the overall model by continuously reducing the prediction error. In each iteration, GBDT gradually reduces the error by building a new decision tree to fit the residual of the current model (i.e. the difference between the predicted value and the true value). The steps are as follows:

Through several iterations to train a new decision tree to fit the residual of the current model, the updated formula is:

$$F_m(x) = F_{m-1}(x) + \alpha h_m(x) \tag{7}$$

Where Fm(x) represents the M-th round model and hm(x) is the newly fitted decision tree.

For the gradients of the loss function there are:

$$g_i^{(m)} = \frac{\partial L(y_i, F_{m-1}(x_i))}{\partial F_{m-1}(x_i)} = -(y - F(x)) \tag{8}$$

Where L is the loss function, we take the square error, Fm-1(x)

For a random forest, the variance of the model is first reduced by outputting the mean value of the prediction results of multiple decision trees:

$$\hat{y} = \frac{1}{M} \sum_{m=1}^{M} T_m(X) \tag{9}$$

Where M is the number of decision trees; $T_m(X)$ is the output of the m-th decision tree; X is the input feature vector, including the type of core material, temperature and excitation waveform. The importance of the feature is:

$$I(X_j) = \sum_{t=1}^{T} \Delta I_p(t, X_j) \tag{10}$$

$I_p$ is the reduced impurity in node t due to feature splitting.

The model is continuously optimized by gradient descent, and the prediction results are obtained by weighting multiple learning trees:

e feature importance index in the above model to quantify the independent influence of each feature on the core loss.

In order to analyze the synergistic influence between two or more features, we introduce the concept of feature interaction: to define the improvement of model performance, that is, the accuracy of prediction results, when two features are split at the same time. The specific formula is as follows:

$$\hat{y}_i = \sum_{m=1}^{M} \lambda_m h_m(X_i) \tag{11}$$

Where $h_m(X_i)$ is the prediction result of the *m* th learning trees; $\lambda_m$ is the learning rate; M is the total number of trees. The model is optimized by learning the residuals step by step, and the residuals for round m are:

$$r_i^{(m)} = y_i - \hat{y}_i^{(m-1)} \tag{12}$$

And in each round, a new learning tree $h_m(X)$ is trained to fit the current residual r.

## 2.3 Analysis of independent and cooperative effects of features

$$\text{We use th} I_n(X_j, X_k) = \sum_{t=1}^{T} \Delta I_m(t, X_j, X_k) \tag{13}$$

According to the results of the previous model, we know that the feature importance of each factor is ranked as the temperature has the greatest influence, the material type has the least influence, and the excitation waveform is between the two. In order to find the conditions under which the core loss is minimized, we use the grid setting search to find the minimum loss under the combination of the three features.

Under the basic model mentioned above, we assume that the core loss model is

$$\begin{aligned} P &= f(N,T,F) \\ &= \alpha_1 N + \alpha T + \alpha F + \epsilon \end{aligned} \tag{14}$$

In the formula, N, T and F respectively represent the type of magnetic core material, temperature and excitation waveform; ε is the noise term; α is the weight of the individual features learned by the model.

**2.4 Model solution and analysis**

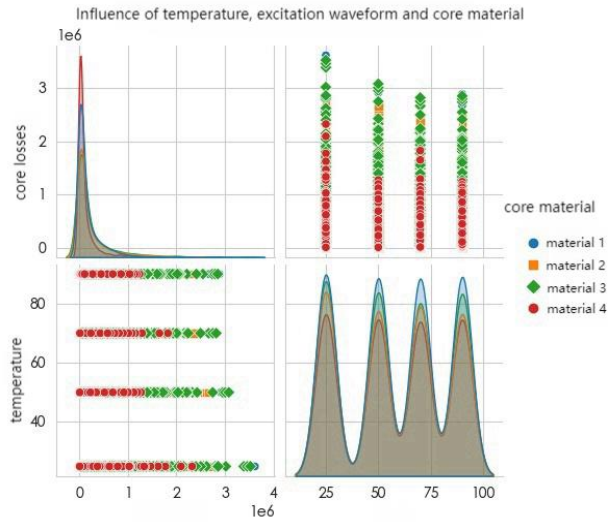

FIG. 1 Distribution and combination diagram of material properties

1. Top left figure (core loss vs core material) : the core loss is concentrated in the lower interval, and the loss of most core materials tends to the area near zero. The loss of materials 1 and 2 is the largest, reaching a peak of about 3e6, while the loss of materials 3 and 4 is lower, and the distribution is more concentrated in a smaller loss range.

2. Top right graph (temperature vs core material) : The core loss distribution of the four materials at different temperatures shows a periodic change in temperature. At temperature nodes such as 25°C, 50°C, 75°C and 100°C, the core loss presents regular peaks, and the distribution of different materials under the same temperature conditions is similar. The loss values of materials 1, 2 and 3 are high, while material 4 is relatively low.

3. Bottom left figure (vspH value of magnetic core loss) : With the change of pH value, the trend of magnetic core loss distribution is not significant, and the loss value of the four materials is basically the same, and concentrated in the lower interval. This means that the pH value has less influence on the core loss.

4. Bottom right graph (temperature vs core material) : Similar to the top right corner, there are periodic fluctuations in the effect of temperature changes on different materials. There are peaks every 25°C, and the loss of the four materials fluctuates as the temperature increases. Material 1 consistently had the highest loss values and material 4 was relatively low.

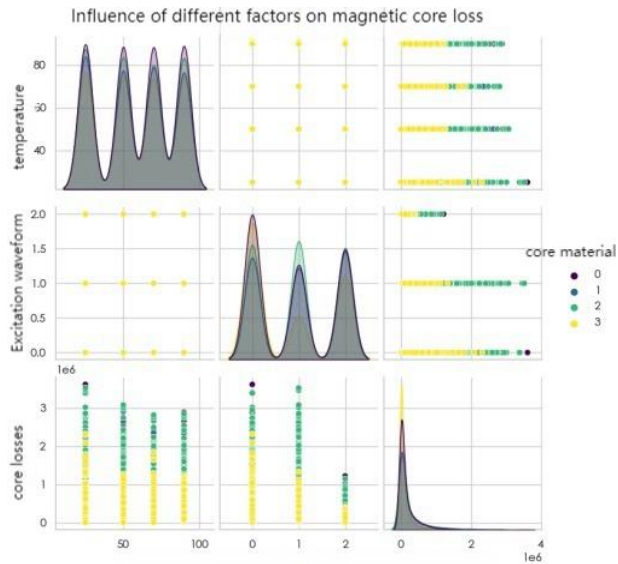

FIG. 2 Summary diagram of each factor affecting magnetic loss

It can be observed that the magnetic core loss of different materials under the same conditions is very different, which indicates that material selection is important to reduce the magnetic core loss.

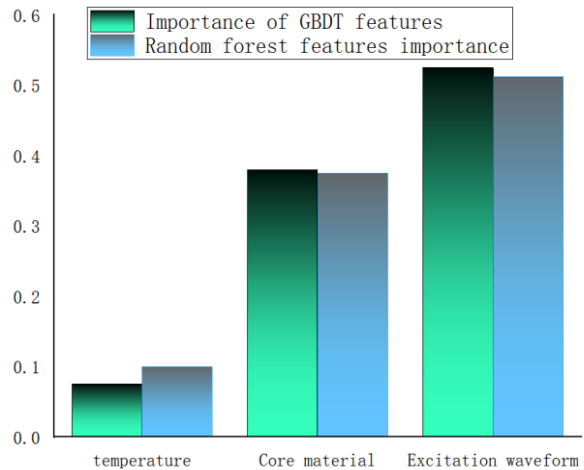

FIG. 3 Comparison of importance between Random Forest and GBDT

Cross-model consistency: Although feature importance rankings may vary across models, two features, the excitation waveform and the core material, show relatively high importance scores across all three models. This may mean that these two features are critical to solving the problem or making accurate predictions.

Feature differences: Temperature features generally have lower importance scores in all three models compared to excitation waveforms and core materials. This suggests that temperature may not be the key factor affecting the predicted results in this problem, or it may be less important than the other two features.

Table 2 Random Forest-GBDT importance comparison table1

| The importance of influencing factors | Random Forest | GBDT |
| --- | --- | --- |
| Core material | 0.1003 | 0.0741 |
| Temperature | 0.3733 | 0.3852 |
| Excitation waveform | 0.1003 | 0.5408 |

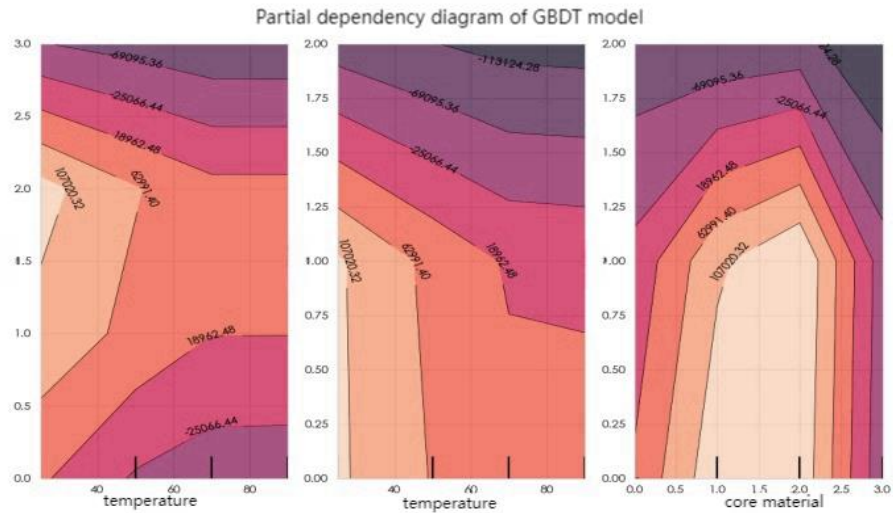

Figure 4 GBDT dependence profile

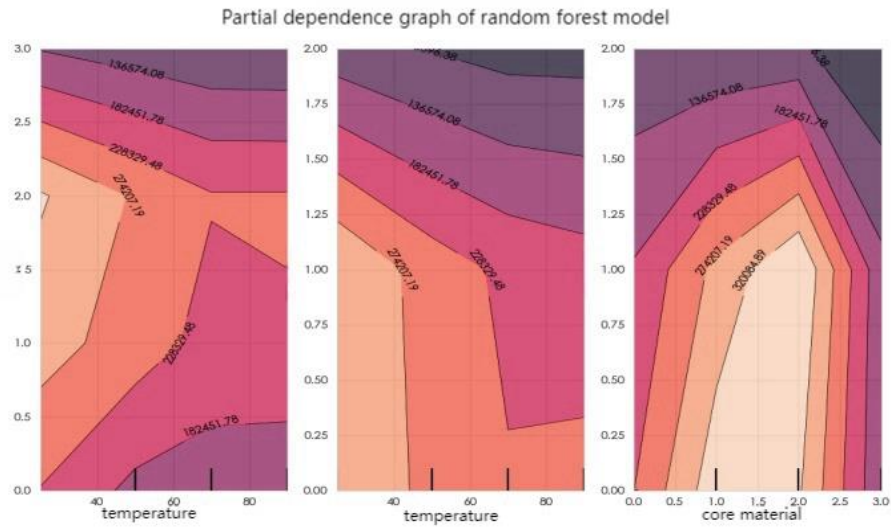

FIG. 5 Dependence distribution map of random forest

1.GBDT model results: excitation waveform (0.5408) has the greatest influence on core loss, indicating that excitation waveform occupies a dominant position in GBDT model; Temperature (0.3852) also has a considerable influence on the core loss, but it is weaker than the excitation waveform; The core material (0.0741) has the least effect on the core loss and contributes little in the GBDT model.

2. Results of random forest model: excitation waveform (0.5264) is still the most important feature in random forest model; Temperature (0.3733) ranked the second, with a greater impact, but not as important as excitation waveform; The magnetic core material (0.1003) also has the least effect, but it is slightly higher than the GBDT model.

3. General trend: excitation waveform is the factor that has the greatest influence

on core loss in the two models; Temperature in the two models of the influence is second, also played a role; The magnetic core material has a relatively small effect and is not the main driver in either model; This suggests that changes in excitation waveform and temperature have a more significant effect on core loss, while the core material has a smaller effect on core loss.

The results are as follows: minimum core loss: 35310.9988; Corresponding conditions: core material =3.0, temperature =80.15 degrees, excitation waveform =2.0.

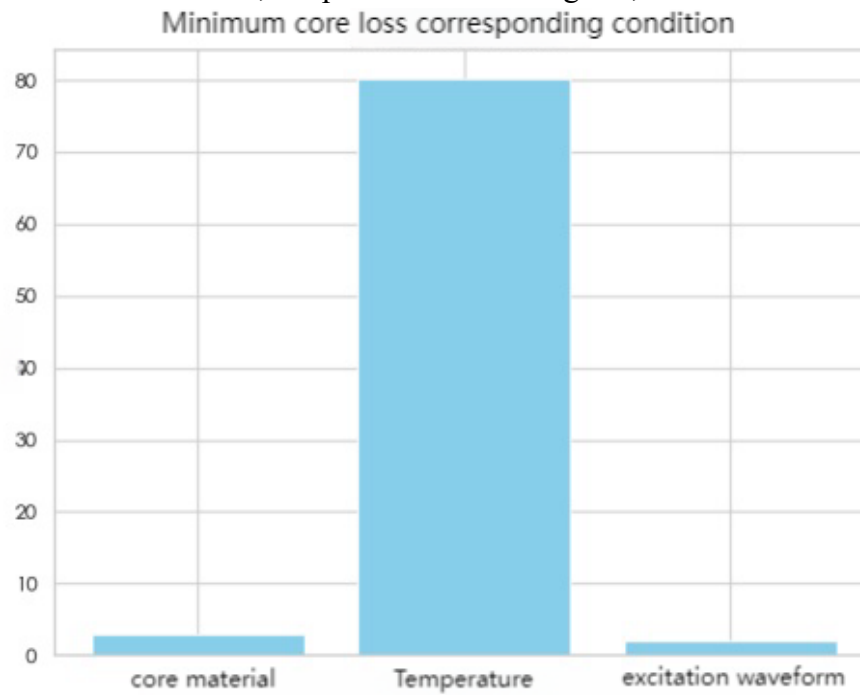

FIG. 6 Minimum core loss related characteristic values

## 3. Core loss prediction

For core loss, we define it as a function of many factors:

$$P = f(N, T, F, f, B_m) \tag{15}$$

Where $B_m$ is the peak magnetic flux density.

For magnetic core loss, we divide it into two main components: hysteresis loss and eddy current loss

For the hysteresis loss there are:

$$P_h = k_h \cdot B_m^n \cdot f^m \tag{16}$$

$P_h$ is the hysteresis loss; $k_h$ is material correlation constant; $B_m$ is magnetic flux density; f is the frequency.

For eddy current loss are:

$$P_e = k_e \cdot B_m^2 \cdot f^2 \cdot d^2 \tag{17}$$

Where d is the thickness of the material.

Therefore, the sum of the core loss is:

$$P = P_h + P_e = k_e \cdot B_m^2 \cdot f^2 \cdot d^2 + k_h \cdot B_m^n \cdot f^m \qquad (18)$$

In order to ensure the accuracy of the model and consider the relevant interference factors, we also introduce a correction factor to the core loss function:

$$P' = (k_e \cdot B_m^2 \cdot f^2 \cdot d^2 + k_h \cdot B_m^n \cdot f^m) \cdot \phi(N,T,F) \qquad (19)$$

For this correction factor is divided into material, temperature, excitation waveform of three factors, respectively expressed as:

$$\begin{cases} \phi_N = c_1 + c_2 \cdot N \\ \phi_T = e^{\alpha(T-T_0)} \\ \phi_F = \beta \cdot F \end{cases} \qquad (20)$$

In summary, the final core loss formula under the introduction of the correction factor is as follows:

$$P_{loss} = (k_e \cdot B_m^2 \cdot f^2 \cdot d^2 + k_h \cdot B_m^n \cdot f^m) \cdot (c_1 + c_2 \cdot N) \cdot e^{\alpha(T-T_0)} \cdot \beta \cdot F \qquad (21)$$

Similar to the previous questions, we also use machine learning methods to fit the model parameters and train the model through the minimum loss function:

$$L = \frac{1}{n} \sum_{i=1}^{n} (y_i - P_{loss}(x_i))^2 \qquad (22)$$

Evaluate the performance of the model by means of mean square error and R square, as follows:

$$\begin{aligned} MSE &= \frac{1}{n} \sum_{i=1}^{n} (y_i - \hat{y}_i)^2 \\ R^2 &= 1 - \frac{\sum_{i=1}^{n}(y_i - \hat{y}_i)^2}{\sum_{i=1}^{n}(y_i - \overline{y_i})^2} \end{aligned} \qquad (23)$$

## 3.1 Model solution and analysis

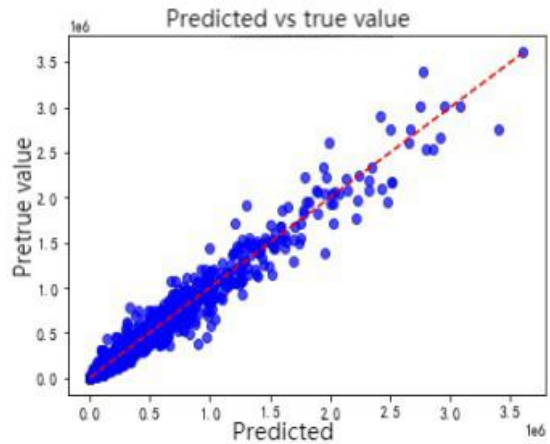
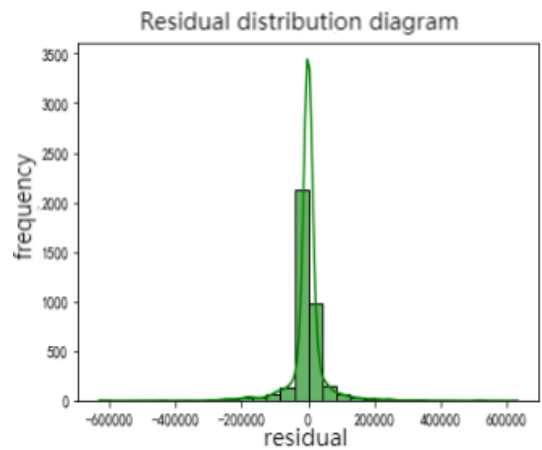

Figure 7 Prediction - real scatter comparison diagram

Figure 8 Residual distribution diagram

Most of the points are close to the ideal diagonal (y=x), indicating that the model's prediction is relatively consistent with the true value 'Overall, the model's prediction is good, with an $R^2$ value of 0.9703, indicating that about 97% of the variation can be explained by the model. The scatter points in the graph are concentrated, showing a small margin of error. The residual distribution is close to the normal distribution, and the center is close to 0, which indicates that the prediction error is relatively uniform and there is no obvious systematic bias. The density curve (KDE) of the residuals shows that most of the residuals are between -100 and 100, indicating that the predicted value differs little from the true value.

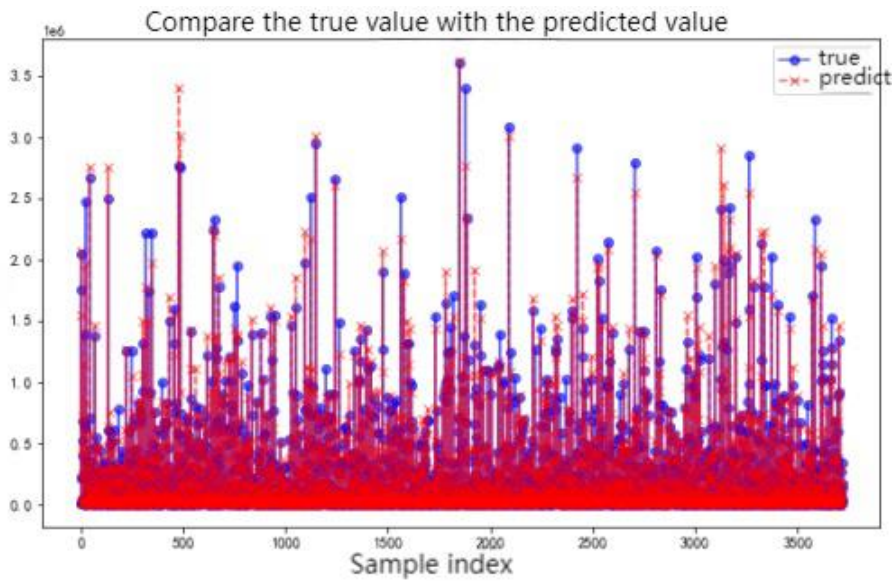

Figure 9 True-predicted value comparison graph

The true value (blue line) and predicted value (red line) maintained a high degree of agreement across most samples and moved very close together, indicating that the model performed well in capturing changes in the data. Although the predicted value deviated slightly from the true value on some samples, the overall trend agreement was good, showing the effectiveness of the model. The visible fluctuations and trend changes were well captured by the model.

Combining $R^2$ values and MAPE(16.7201), the model performs well in predicting core loss with relative accuracy. Although there is some absolute error, it can be considered that the model has certain reliability and practicability in practical application through the analysis of each graph. It is recommended to continue to optimize the model to reduce the errors and improve the prediction accuracy.

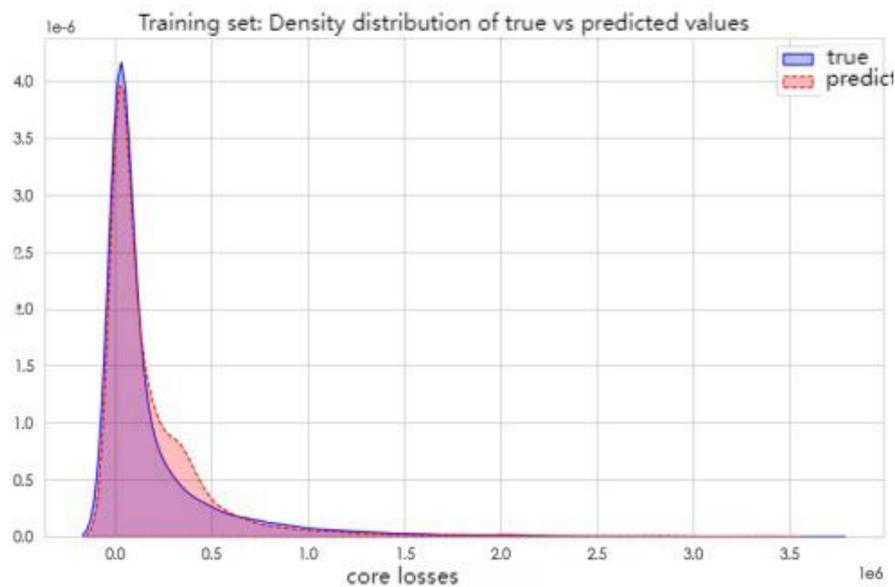

Figure 10 Density distribution of training set

For the variable core loss (unit: w/m3), the density distribution of the true value and the predicted value in the training set. As can be seen from the figure, the density distribution of the real value shows a trend of decreasing first and then increasing, specifically from 4.0, gradually decreasing to about 3.5, and then continuously decreasing to about 2.0. This indicates that in the real case, the numerical distribution of magnetic core loss has a changing process from high to low.

However, the density distribution of the predicted values showed a different trend. The predicted value starts at around 3.5 and drops to around 2.0, but then gradually rises to around 3.5. This trend is not completely consistent with the downward trend of the real value, especially in the lower core loss interval (below 2.0) and higher interval (above 3.5), the difference between the predicted value and the real value is more obvious.

On the whole, although the predicted value captures the trend of the real value to a certain extent, the prediction accuracy in some areas (such as the extreme value region of the core loss) may not be high enough.

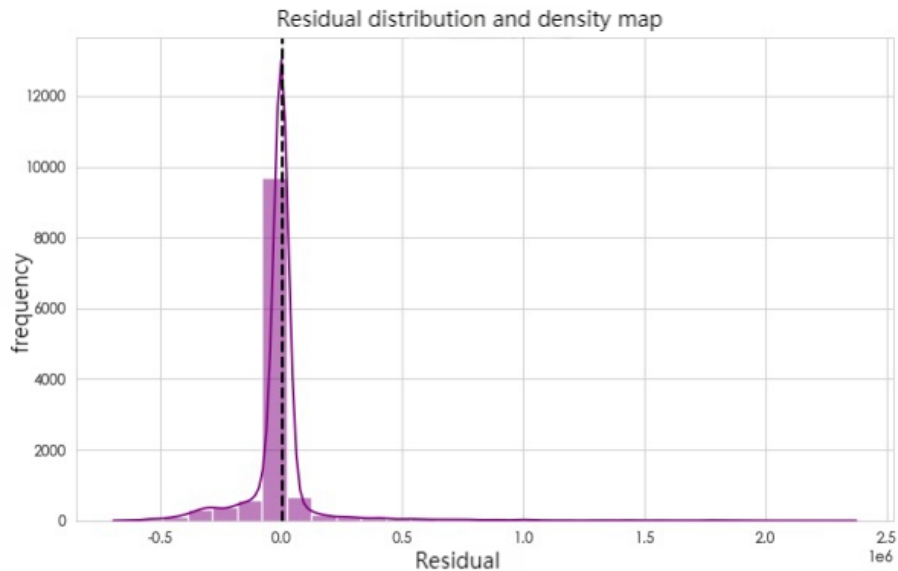

Figure 11 Residual distribution diagram

This graph shows the residual distribution of random variables and the corresponding density situation. It is clear from the graph that the residual distribution has a striking feature of a very sharp peak at 0.5. This spike indicates a high concentration of data points at around 0.5 residuals, and may mean that there are some outliers or extremes in the data that cause the residuals at that location to be abnormally high.

In addition, the entire distribution presents a significant skew. Specifically, in the negative region of the residuals (i.e., the left side), the density is higher, indicating that the residuals tend to shift in the negative direction; While in the positive region of the residuals (i.e. the right), the density gradually decreases, especially after greater than 1.0, the density decreases significantly. This skew indicates that the residuals in the data are not perfectly symmetrical, but rather have a certain directionality.

## 4. Multi-objective optimization model

In order to build an optimization model that considers both core loss and transmitted magnetic energy, we can introduce a multi-objective optimization method, the specific steps are as follows:

### 4.1 Construction of multi-objective optimization model

**1. Definition of relevant variables**

Based on the above related parameters, we summarized the parameters as follows:

$$\begin{cases} T : \text{temperature} \\ f : \text{frequency} \\ B : \text{Peak magnetic flux density} \\ N : \text{Type of magnetic core material} \\ F : \text{Excitation waveform} \\ CL : \text{core losses} \\ ME : \text{Transmitted magnetic energy} \end{cases} \quad (24)$$

**2. Establishment of optimization objectives**

For the core loss

$$CL = (T, f, B, N, F) \quad (25)$$

Our optimization goal is to minimize the core loss CL and maximize the transmitted magnetic energy ME, so there are:

$$\begin{cases} Min\ CL \\ Max\ ME \end{cases} \quad (26)$$

In order to simplify the model, we define a comprehensive function to convert the solving problem into a single objective optimization problem, introduce the weight coefficient, and judge the optimization result by adding or subtracting the two objective values of a certain proportion, namely:

$$Min\ Z = \omega_1 CL - \omega_2 ME \quad (27)$$

**3. Constraints under actual working conditions**

Considering that there will be certain restrictions on temperature, magnetic flux density and frequency under actual working conditions, that is:

$$S.t \begin{cases} T_{min} \leq T \leq T_{max} \\ f_{min} \leq f \leq f_{max} \\ B_{min} \leq B \leq B_{max} \end{cases} \quad (28)$$

Apply a certain feasible range to the peak temperature, frequency and magnetic flux density to ensure the rationality of the results.

**4.2 Find the optimal solution based on genetic algorithm**

Genetic algorithms seek solutions to optimization problems by simulating the process of biological evolution. The core idea is to make individuals in a population evolve gradually and tend to find the optimal solution to the problem through operations such as selection, crossover and variation. Here's how to solve the single-objective optimization above by using genetic algorithms:

- **Target definition and initialization**

For a single objective optimization problem, we need to find a set of variables X that minimizes the objective function Z, namely:

$$Min\ Z = \omega_1 CL - \omega_2 ME \tag{29}$$

By initializing each individual of the population, randomly generated within the range of various variable parameters:

$$\mathbf{x}_i = \mathbf{x}_{\min} + \text{rand}() \times (\mathbf{x}_{\max} - \mathbf{x}_{\min}), \quad i = 1, 2, ..., N \tag{30}$$

- **Selection of fitness function**

Evaluate the fitness of each individual, the higher the fitness of the individual has a greater chance of being selected for breeding. Since it is solving for the minimization objective function, there is

$$\text{fitness}(\mathbf{x}) = \frac{1}{1 + f(\mathbf{x})} \tag{31}$$

To select individuals based on fitness calculation in the above formula and select individuals with higher fitness with a higher probability, we take roulette selection rule as follows:

$$P_i = \frac{\text{fitness}(\mathbf{x}_i)}{\sum_{j=1}^{N} \text{fitness}(\mathbf{x}_j)} \tag{32}$$

- **Crossover and variation**

In order to exchange some genes of the selected parent to generate new offspring and introduce diversity, suppose the parent is x1, x2, and the crossing point is k, then:

$$\begin{aligned} \mathbf{y}_1 &= (x_{11}, x_{12}, ..., x_{1k}, x_{2(k+1)}, ..., x_{2n}) \\ \mathbf{y}_2 &= (x_{21}, x_{22}, ..., x_{2k}, x_{1(k+1)}, ..., x_{1n}) \end{aligned} \tag{33}$$

At the same time, in order to prevent local optimality, we take mutation operation to further increase the population diversity, and by randomly selecting some genes of the individual to carry out small-scale perturbation, there are:

$$x_j' = x_j + \delta \tag{34}$$

j is the mutation point and the perturbation term. $\delta$

- **Generation and termination**

The population composed of new individuals obtained through the above selection, crossover, mutation and other operations can carry out the next evolution, and the termination condition of the program can be specified as reaching the maximum number of iterations.

## 4.3 Analysis of model

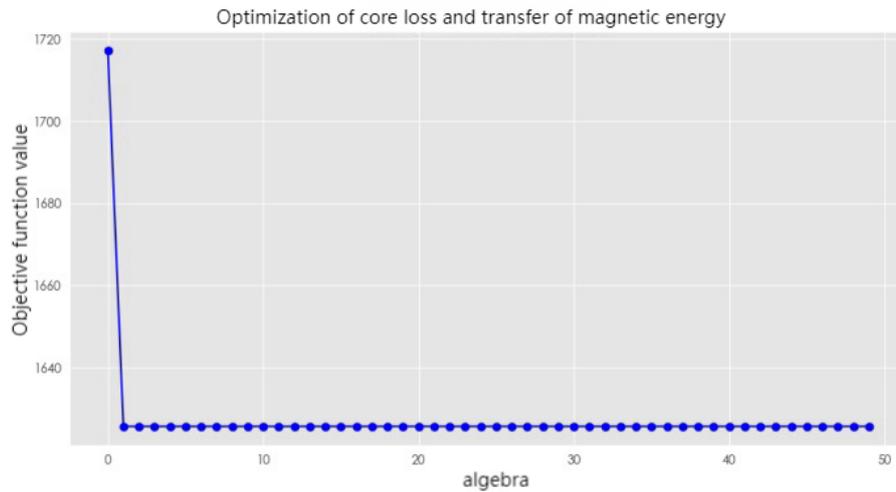

FIG. 12 Optimization of core loss and transferred magnetic energy

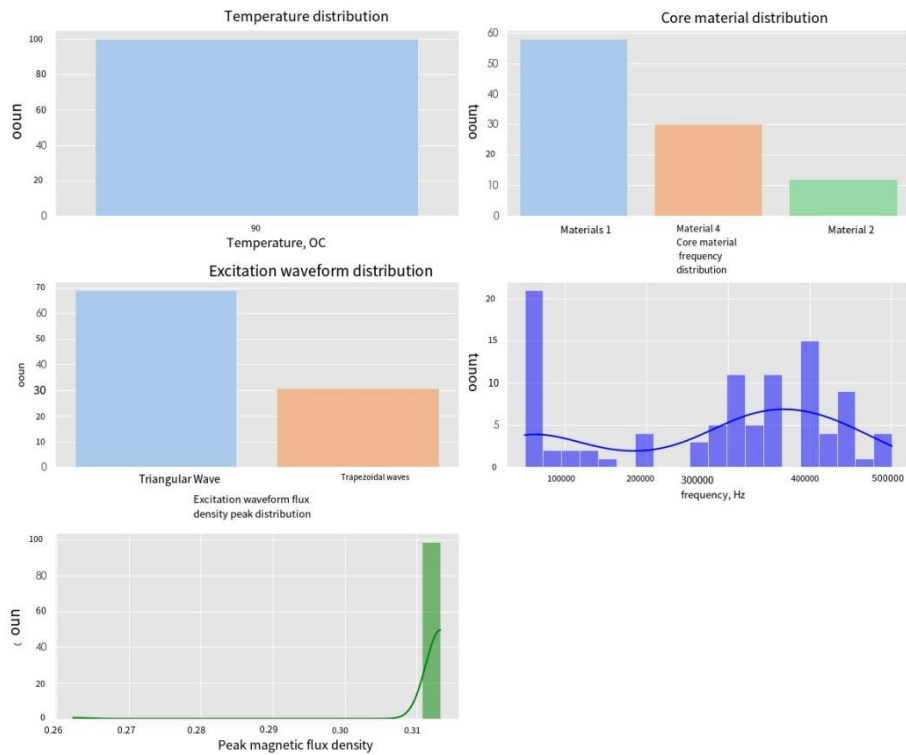

FIG. 13 Distribution of related parameters

The distribution of temperature, material and excitation waveform shows the selection preferences of these parameters. It can be observed that at a temperature of 90°C, there may be more individual selection, reflecting the trend of the optimization process; The distribution of core materials and excitation waveforms can reveal the popularity of particular materials and waveforms in the population, indicating the effectiveness of these combinations in reducing losses.

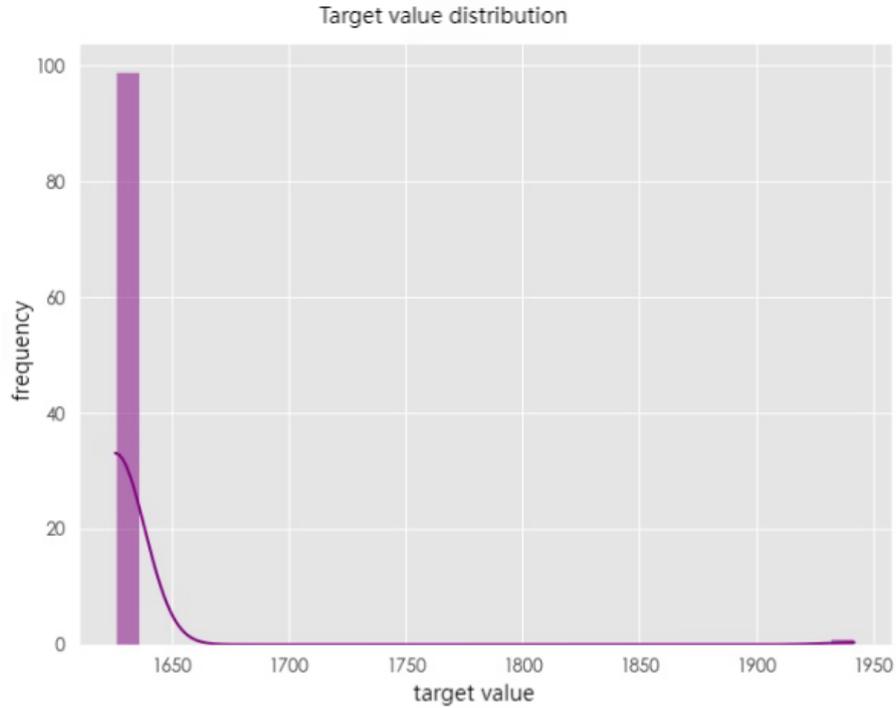

Figure 14 Target value distribution diagram

The histogram of the target value shows the target value distribution of individual individuals in the population, with higher frequencies concentrated in the lower target value region, indicating that the vast majority of individuals have a lower loss level after optimization. The KDE curve (kernel density estimation) clearly depicts the trend of target value distribution, which can help to identify the possible optimal value interval. The target value of the best combination is: 1625.69854; The core loss of the best combination is: 509.30773; The transmission magnetic energy when the best combination is: 0.313284.

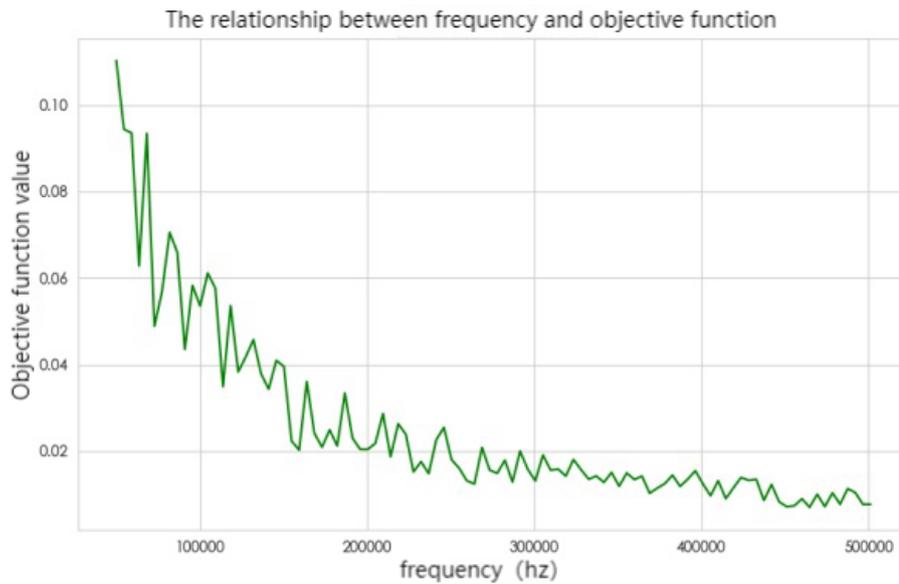

Figure 15 Frequency relationship diagram1

It can be seen from the figure above that as the frequency increases from 0 to

500000Hz, the value of the objective function presents a certain trend of change. The value of the objective function also fluctuates or oscillates during the process of frequency increase. This means that at some frequency points, the value of the objective function may rise, while at others it falls, forming one or more peaks and troughs; In addition to the simple monotone or fluctuating trend described above, the objective function value also exhibits a more complex non-linear change as the frequency increases. This change involves multiple phases, each with a different tendency to increase or decrease.

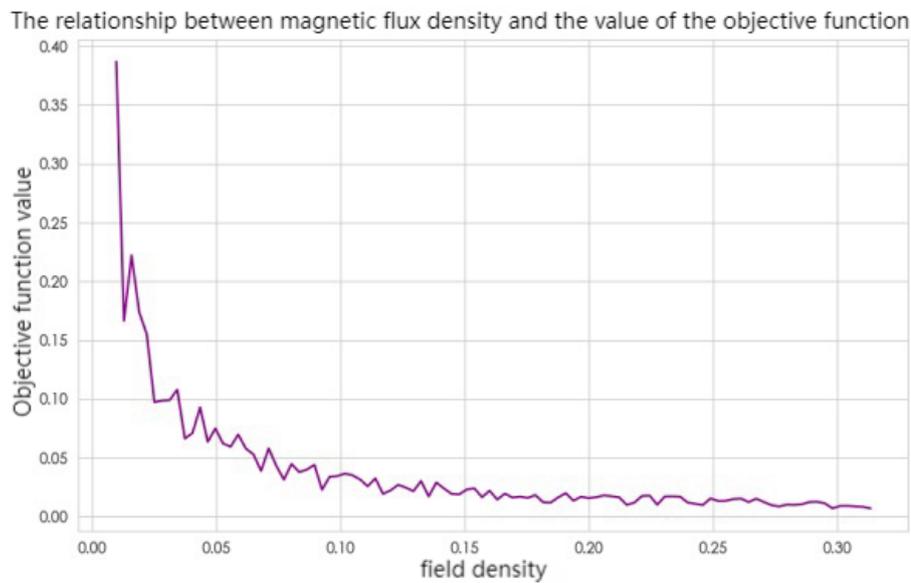

Figure 16 Line chart of magnetic flux density relationship

The value of the objective function begins to fluctuate or oscillate. This is due to factors such as resonant effects within the system, nonlinear interactions, or external interference. Fluctuations and oscillations appear as periodic ups and downs in the value of the objective function as it increases or decreases. As the magnetic flux density approaches or reaches its maximum (0.30 Tesla), the objective function value may gradually stabilize.

Table 3 Summary table of results2

| Optimal conditions | Results |
| --- | --- |
| Temperature (ºC) | 25 |
| Core material | Material 2 |
| Excitation waveform | Sine wave |
| Frequency (Hz) | 493661.35 |
| Peak magnetic flux density | 0.3126 |
| Optimal objective function value | 0.00657 |

## 4.4   Design of Core Compensation for 75kW Tidal Current Energy Generator

A thorough analysis of the mechanism and influencing factors of core losses in magnetic cores has laid a solid theoretical and technical foundation for solving the core losses problems of various electrical equipment. We designed a 75kW tidal energy generator and optimized its structure based on the magnetic core compensation model. The actual driving test of this motor performed well.

By deeply analyzing the operating conditions and performance requirements of the 75kW tidal energy generator and applying the models and rules constructed during the previous analysis of core losses in a targeted manner, the selection of core material, structural design, and operating parameters of the generator can be optimized, thereby effectively reducing core losses and improving the overall performance of the generator.

Electromagnetic topology selection: The 75kW permanent magnet motor adopts a compact permanent magnet motor topology, which belongs to a radial magnetic field outer rotor structure.

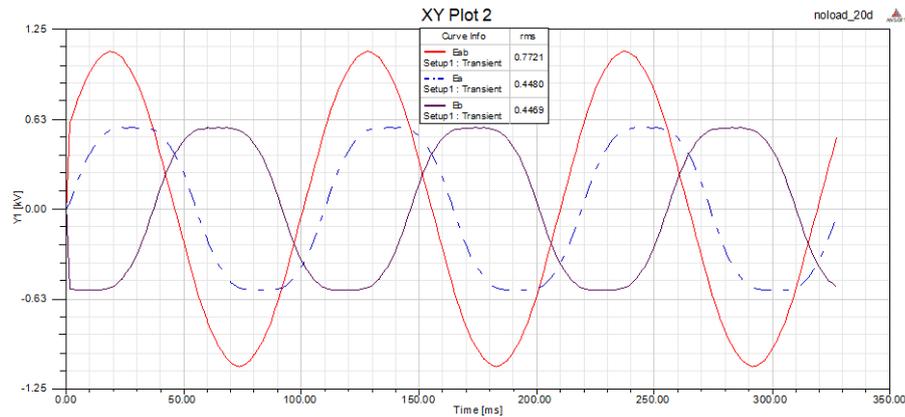

Figure 17 No-load reverse electromotive force

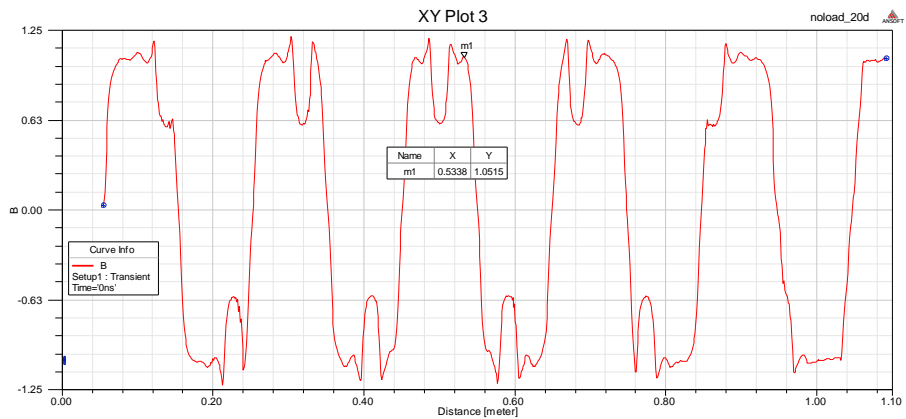

Figure 18 Air-gap magnetic flux density without load

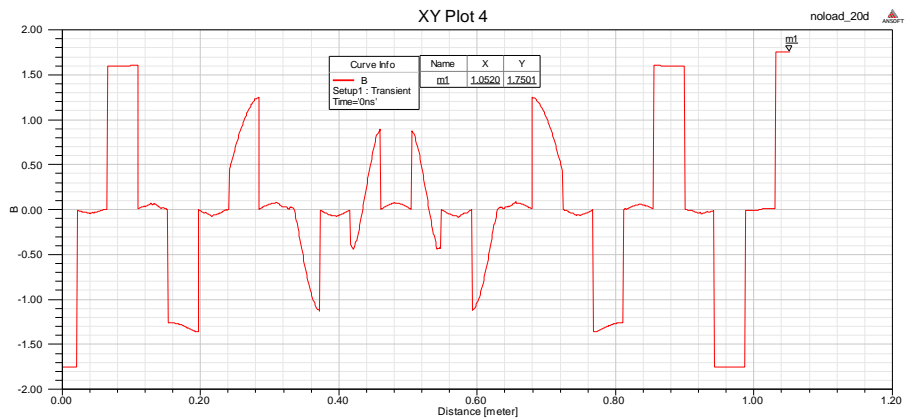

Figure 19 No-load tooth magnetic density

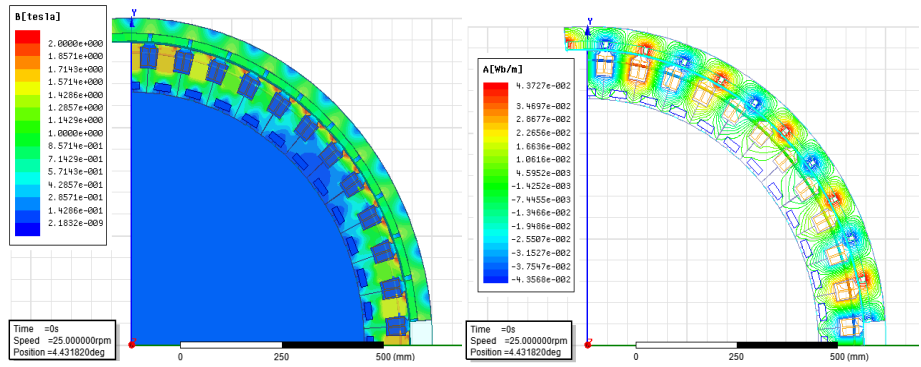

Figure 20 Magnetic density distribution

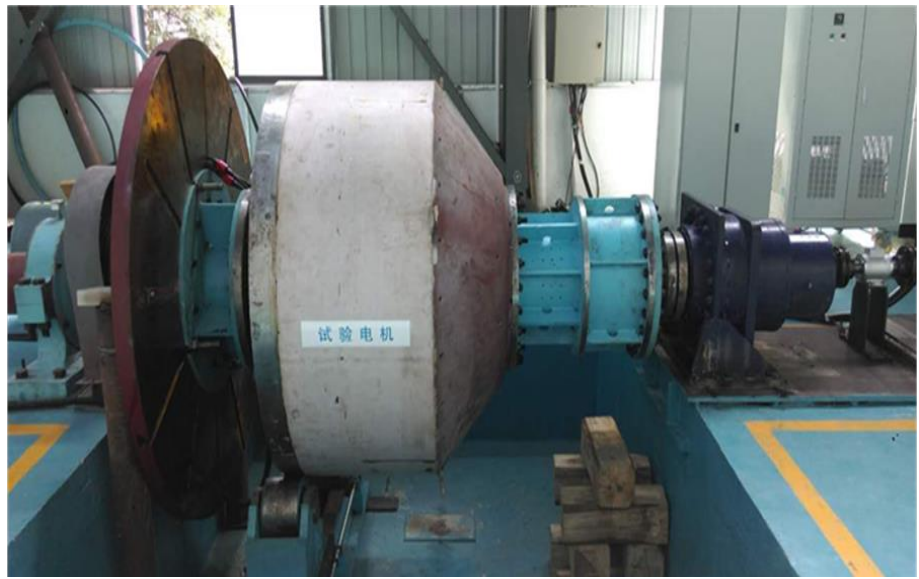

Figure 21 Experimental motor diagram

In terms of losses, through the analysis of stator copper loss, rotor iron loss and permanent magnet eddy current loss, it was found that copper loss accounted for a relatively large proportion in the total loss. This provides a direction for subsequent optimization. By optimizing the design of stator windings and reducing the winding resistance, the copper loss can be further decreased. In addition, by adopting reasonable partition and shielding measures for the permanent magnet, the eddy current loss of the permanent magnet can be effectively reduced, and the efficiency of the motor can be improved.

Overall, the radial field external rotor topology structure of the 75kW compact permanent magnet motor has shown excellent performance in output performance, operational stability and energy efficiency, proving the applicability of this topology structure in tidal energy generation. However, there is still some room for optimization. By making targeted improvements to the losses of each part of the motor, the overall performance of the motor can be further enhanced.

## 5. Summary

During the research on the magnetic core of the 75kW tidal current generator, key issues such as classification of excitation waves, correction of the Steinmetz equation, analysis of factors influencing magnetic core loss, data-driven magnetic core loss prediction model, and determination of the optimal conditions for magnetic components were addressed. The research team utilized data-driven analysis methods, integrating machine learning and optimization algorithms, to build a series of models for analyzing and optimizing the design of magnetic components. The multi-objective optimization algorithm played a significant role in minimizing the magnetic core loss of the tidal current generator and maximizing the transmission of magnetic energy. After magnetic core loss correction and optimized design, the 75kW tidal current generator met the design requirements during the driving test and performed well in operation. Looking forward to the future, in order to meet the complex design requirements of the magnetic core of the tidal current generator, it is necessary to deeply integrate physical models with data-driven methods, enhance the interpretability, computational efficiency, and prediction accuracy of the models, and continuously promote the progress of tidal current power generation technology.